\documentclass[12pt]{article}

\usepackage{graphicx,amsmath,amssymb,color,bm}

\begin{document}

\centerline {\Large\textbf {The diverse magneto-optical selection rules
}}

\centerline {\Large \textbf {in bilayer black phosphorus}}\vskip0.6 truecm

\centerline{Jhao-Ying Wu$^{1,*}$, Szu-Chao Chen$^{2,\dag}$, Thi-Nga Do$^{3}$, Wu-Pei Su$^{4}$, and Godfrey Gumbs$^{5,6\dag\dag}$}

\centerline{$^{1}$Center of General Studies, National Kaohsiung Marine University, Kaohsiung, Taiwan 811}
\centerline{$^{2}$Department of Physics, National Cheng Kung University, Tainan, Taiwan 701}
\centerline{$^{3}$Department of Physics National Kaohsiung Normal University}
\centerline{$^{4}$Department of Physics, University of Houston, Houston, Texas}
\centerline{$^{5}$Department of Physics and Astronomy, Hunter College at the City University of New York,\\ \small }
\centerline{695 Park Avenue, New York, New York 10065, USA}
\centerline{$^{6}$Donostia International Physics Center (DIPC),}
\centerline{Paseo de Manuel Lardizabal 4, 20018 San Sebasti\'an Donostia, Spain}

\begin{abstract}

The magneto-optical properties of bilayer phosphorene is investigated by the generalized tight-binding model and the gradient approximation. The vertical inter-Landau-level transitions, being sensitive to the polarization directions, are mainly determined by the spatial symmetries of sub-envelope functions on the distinct sublattices. The anisotropic excitations strongly depend on  the electric and magnetic fields.
A perpendicular uniform electric field could greatly diversify the selection rule, frequency, intensity, number and form of symmetric absorption peaks. Specifically, the unusual magneto-optical properties appear beyond the critical field as a result of two subgroups of  Landau levels with the main and side modes. The rich and unique magneto-absorption spectra arise from the very close relations among the geometric structures, multiple  intralayer and interlayer hopping integrals, and composite external fields.

\vskip0.6 truecm

\noindent

\end{abstract}

\newpage

\bigskip

\centerline {\textbf {I. INTRODUCTION}}%

Black phosphorus (BP), which is the most stable allotrope of the group-V element phosphorus, are successfully synthesized under the few-layer structure referred to as phosphorene. The various experimental techniques to obtain the novel material include the use of mechanical cleavage \cite{PLi2014,PLiu2014}, liquid exfoliation \cite{PBrent2014,PYasaei2015,Pkang2015}, as well as mineralizer-assisted short-way transport reaction \cite{PLange2007,PNilges2008,PKopf2014}. Phosphorene presents an energy gap in the range of ${\sim\,0.5-2}$ eV \cite{gap1,PRudenko,gap} sensitive to the number of layers, as verified by optical experiments \cite{PLiu2014,Zhang2014}. Such band gaps are larger than that of bulk counterpart (${\sim\,0.3}$ eV) \cite{gap1,Li2014,Han2014}, and are in sharp constrast with the zero or narrow gaps of two-dimensional (2D) group-IV materials \cite{Balendhran2015}. Transport measurements have demonstrated that the phosphorene-based field-effect transistor exhibits an on/off ratio of $10^{5}$ and the carrier mobility at room temperature as high as $10^{3}$ cm$^{2}/$V$\cdot$s \cite{Li2014}. The high performance implies potential applications in nanoelectronics. Specifically, many remarkable anisotropic physical phenomena are reported, such as mechanical strains \cite{Rodin2014}, excitonic effects \cite{Oleg}, optical spectra \cite{Li2014}; transport \cite{TLow} and thermal \cite{LingX2015} properties. The peculiar characteristics can be traced back to its feature-rich lattice structure and electronic properties.



Phosphorene possesses a puckered structure, being closely related to the $sp^{3}$ hybridization of four $(3s,3p_{x},3p_{y},3p_{z})$ orbitals. The deformed hexagonal lattice in the x-y plane is quite different from the honeycomb lattice of group-IV systems \cite{PRudenko}. This unique geometric structure fully dominates the low-lying energy bands which are highly anisotropic in the dispersion relations as a function of the in-plane wave vector ${\bf k}$, i.e., the linear and parabolic dispersions near the Fermi energy $E_{F}$, respectively, along the $\widehat{k_{x}}$ and $\widehat{k_{y}}$ directions \cite{PRudenko}. The carrier effective masses are anisotropic and asymmetric between electrons and holes \cite{Oleg,Qiao2014}. Accordingly, the best functional performances in  nanoelectronics and optoelectronics depend on the rotation-induced orientations of BP systems, which could be deduced from the optical spectroscopy measurements \cite{Qiao2014}. The strong anisotropy sets phosphorene aside from most of 2D materials, such as graphene, boron nitride and transition metal dichalcogenides.




A uniform electric field (${E_z\hat z}$) could greatly diversify the essential properties. In a few-layer BP, it will create a gapless band structure after reaching a critical electric field ($E_{z,c}$) \cite{PDolui,PLiu2015}. The various energy bands come to exist during the variation of $E_z$, including parabolic bands, graphene-like  Dirac cones and  oscillatory energy bands. This arises from a competitive or cooperative relation between the intralayer and  interlayer atomic interactions and the Coulomb potential energy. The strong electrically tunable energy bands enrich the quantization phenomena under a uniform perpendicular magnetic field (${B_z\hat z}$), such as two subgroups of Landau levels (LLs), uniform and non-uniform LL energy spacings, and frequent crossings and anti-crossings \cite{Wu2017}. There exist dramatic changes in the sub-envelope functions for the two mixed LLs with anti-crossing behaviors. They are expected to be responsible for the significant physical properties, e.g., the magneto-optical selection rules.

Phosphorene exhibits the unusual optical properties, such as, the high optical absorption in the UV region and the dichroism. The threshold absorption structure, corresponding to the band-gap energy, is revealed in the light polarization along the armchair direction \cite{Qiao2014}. Its frequency falls rapidly with the number of layers. In contrast, the band-gap absorption is forbidden under the zigzag-direction electric polarization, mainly owing to the specific symmetry of wave functions in the initial and final states. Theoretically speaking, the electric dipole excitations, which connect the valence and conduction band states, are thoroughly different for two perpendicular polarization directions \cite{Qiao2014}. The strong anisotropy is also predicted in the magneto-optical conductivity of BP thin films \cite{Zhou2015}.


In this work, we investigate the tuning effects, brought about by a perpendicular electric field, on the magneto-optical absorption spectra of bilayer phosphorene. The relations among the geometric structure, intrinsic interactions and external fields are explored in detail. The generalized tight-binding model \cite{CYLin2015} has been developed to investigate the electronic properties of monolayer and bilayer phosphorus in a composite electric and magnetic field. The main features include the $E_z$-induced dramatic changes in energy bands and LLs \cite{Wu2017}. The diverse magnetic quantizations will be directly reflected in the magneto-optical excitations. The electric-dipole transition elements are evaluated within the gradient approximation \cite{Pedersen2011,Huang2008}. The magneto-optical absorption spectra are predicted to strongly depend on the polarization directions and the electric- and magnetic-field strengths. By the delicate calculations and the detailed analyses, the available inter-LL transitions are deduced to be mainly determined by the spatial distributions/symmetries of sub-envelope functions on the distinct sublattices. The selection rule, frequency, intensity, number and structure of absorption peaks are dramatically changed by the $E_z$ field. Whether the usual/unusual optical properties are revealed in magneto-absorption spectra relies on the electric field beyond the critical one or nor. The predicted results could be further validated by the magneto-optical measurements \cite{Chuang2009,Orlita2009,Ubrig2011,Plochocka2012,Plochocka2008,Orlita2015,Zaric2004,Berciaud2014,Henni2016}.




\bigskip
\bigskip
\centerline {\textbf {II. METHODS}}%
\bigskip
\bigskip

Monolayer phosphorene, with a puckered honeycomb structure, has a primitive unit cell covering four P atoms, as clearly shown by the dashed green lines in Fig.\ 1(a). Two of them are located on the lower (red circles) or higher (blue circles) sublattice sites; that is, there exists four sublattices (${A_1, A_2, B_1, B_2}$). Similar structures are revealed in few-layer systems, e.g, bilayer phosphorene in Fig.\ 1(b). The low-lying energy bands are dominated by the atomic interactions of 3$p_{z}$ orbitals \cite{PRudenko}. The few-layer Hamiltonian is characterized by

\begin{equation}
H=\sum_{l}\sum_{i=1}^{4}(\varepsilon_{i}^{l}+U_{i}^{l})c_{i}^{l}c_{i}^{\dag\,l}+
\sum_{l,\langle i,j \rangle}h^{ll}_{ij}c_{i}^{l}c_{j}^{\dag\,l}+
\sum_{l\neq l^{\prime},\langle i,j \rangle }h^{\prime\,ll^\prime}_{ij}c_{i}^{l}c_{j}^{\dag\,l^{\prime}} \  .
\end{equation}
$\varepsilon_{i}^{l}$ is zero in a monolayer BP; for a few-layer system, it is a layer- and sublattice-dependent site energy due to the chemical environment difference, e.g., ${\varepsilon^l_i\,=1}$ eV between the A and B sublattices in bilayer BP \cite{PRudenko}. $U_{i}^{l}$ is the Coulomb potential energy induced by an electric field. They both contribute to the diagonal matrix elements. The summation is made on four lattice sites and various layers. $c_{i}^{l}$ ($c^{\dag\,l^{\prime}}_{j}$) is to destroy (create)  an electronic state on the $i$-th ($j$-th) lattice site of the $l$-th (${l^\prime}$-th) layer, where ${i=1, 2, 3; 4}$, respectively, correspond to the P atoms on the $A_1$, $B_1$, $B_2$ and $A_2$ sublattices (Fig. 1(a)). $h^{ll}_{ij}$ and $h^{\prime\,ll^\prime}_{ij}$ are, respectively, the intralayer and interlayer hopping integrals. The effective atomic interactions used in the calculations cover $h_{12}^{11}$=3.665 eV, $h_{21}^{11}=-0.055$
eV, $h_{13}^{11}=-0.105$ eV, $h_{14}^{11}=-0.205$ eV, $h_{41}^{11}=-1.22$
eV, $h_{12}^{21}=0.295$ eV, $h_{21}^{21}=-0.091$
eV, $h_{24}^{21}=-0.151$ eV, and $h_{42}^{21}=0.273$ eV \cite{PRudenko}, as indicated in Figs. 1(a) and 1(b).

Monolayer and bilayer BP systems are present in a uniform perpendicular magnetic field. The magnetic flux through a unit rectangle is $\Phi= a_{1}a_{2}B_z$, where $a_{1}=3.27$ ${\AA}$ and $a_{2}=4.43$ ${\AA}$ are lattice constants (the lattice vectors shown by the green arrows in Fig. 1(a)). The vector potential, $\vec{A}=(B_z x)\hat{y}$, can create the Peierls phase of $exp\{i[\frac{2\pi}{\phi_{0}}\int \vec{A}\cdot d\vec{r}] \}$ in the hopping integrals, leading to a new period along $\hat x$ and thus an enlarged rectangular unit cell with ${4R_B= 4\phi_0\,/\Phi}$ atoms in a monolayer BP, as illustrated in Fig. 1(c). $\phi_{0}$ (=hc/e=4.1$\times 10^{-15}$ T${\cdot}$m$^2$) is the magnetic flux quantum; ${\phi_0\,/\Phi}$ is chosen to be an integer for a model study. The reduced first Brilloun zone has an area of ${4\pi^2\,/a_1a_2R_B}$, in which all the LLs are degenerate for any ${(k_x,k_y)}$ states. For bilayer BP, the magnetic Hamiltonian matrix, which is built from the  ${8R_B}$ tight-binding functions, has ${8R_B\times\,8R_B}$ elements. This Hermitian matrix is very huge within achievable experimental field strengths, e.g., the dimension of 16800 at ${B_z=30}$ T.

After the diagonalization of bilayer magnetic Hamiltonian, the LL wave function, with quantum number $n$, could be expressed as
\begin{equation}
\psi(n,\mathbf{k})=\sum_{l=1,2}\sum_{\alpha=1,2}\sum_{\beta=1}^{R_{B}}[A^{l}_{\alpha,\beta}(n,\mathbf{k})| \phi^{l}_{\alpha,\beta}(A)\rangle+B^{l}_{\alpha,\beta}(n,\mathbf{k})| \phi^{l}_{\alpha,\beta}(B)\rangle],
\end{equation}
where ${\phi^l_{\alpha\,,\beta}}$ is the tight-binding function localized at the sublattice-dependent lattice sites. ${A^l_{\alpha\,,\beta}(n\,,{\bf k}\,)}$/${B^l_{\alpha\,,\beta}(n\,,{\bf k}\,)}$ is the amplitude on the $({A_1^l, A_2^l, B_1^l, B_2^l})$-sublattice-dependent lattice site. Specifically, all the amplitudes in an enlarged unit cell could be regarded as the spatial distributions of the sub-envelope functions on the distinct sublattices. Bilayer BP has eight magnetic sub-envelope functions, in which the dominating ones are utilized to characterize the magnetic quantum numbers and the types of LLs. They provide much information for explaining the peculiar LL behaviors, as discussed below. To achieve the experimentally attainable field strengths, a band-like matrix is developed to solve the huge Hamiltonian matrix efficiently \cite{CYLin2015}. The strong electrically-tunable LL energy spectra exist in $B_{z}<60$ T. The theoretical framework incorporates the intra-layer and inter-layer atomic interactions, as well as the effects due to geometric structures and external fields, simultaneously. It is suitable for studying the diverse quantization phenomena in layered materials with distinct stacking configurations under uniform/modulated/composite external fields. Moreover, the calculated results are accurate and reliable over a wide energy range.

The electric-dipole optical excitations directly reflect the main characteristics of the electronic properties. The occupied valence LL of $n^v$ is vertically excited to the unoccupied conduction LL of $n^v$ in the presence of  an electro-magnetic wave at zero temperature, being characterized by the excitation channel of $\Delta\,n=|n^v-n^c|$. Based on the linear-response Kubo theory, the optical spectral function, without the change of wave vector in the inter-LL transition, is given by
\begin{equation}
A(\omega)\propto 4({4\pi^2\,/a_1a_2R_B})\sum_{n^{c},n^{v}}|\langle \psi^{c}(n^{c},\mathbf{k})|\frac{\mathbf{\hat{E}}\cdot \mathbf{p}}{m}|\psi^{v}(n^{v},\mathbf{k})\rangle|^{2}\times \frac{\Gamma}{[E^{c}(n^{c},\mathbf{k})-E^{v}(n^{v},\mathbf{k})-\omega]^{2}+\Gamma^{2}}.
\end{equation}

The contributions, which come from the spin, localization-center and ${\bf k}$ degeneracies, are included in the first term. The available transition channels and the  excitation intensity depend on the square of the velocity matrix element ${|M(n^c,n^v)|^2}$. The third term is the joint density of states related to the initial [(${n^v,{\bf k}}$)] and final [(${n^c,{\bf k}}$)] LLs. ${\bf {\hat E}}$ is an unit vector of electric polarization, and ${\bf p}$ is momentum. The parallel and  perpendicular polarizations, ${\bf{\hat E}\parallel\,{\hat x}}$ and ${\bf{\hat E}\bot\,{\hat x}}$, are taken into account, respectively. The velocity matrix element $M(n^{c},n^{v})$ is calculated within the gradient approximation and is expressed by the amplitudes $A^{l}_{\alpha,\beta}$'s($B^{l}_{\alpha,\beta}$'s) of the tight-binding functions $(\phi^{l}_{\alpha,\beta})$'s on the distinct sublattices. For ${\bf {\hat E}\parallel\,{\hat x}}$, the velocity matrix element is evaluated from

\begin{eqnarray}
M(n^{c},n^{v})=\sum_{l,l^{\prime}=1,2}\sum_{\alpha,\alpha^{\prime}=1,2}\sum_{\beta,\beta^{\prime}=1}^{R_{B}}\{A^{\ast l^{\prime}}_{\alpha^{\prime},\beta^{\prime}}(n^{v},\mathbf{k})\times A^{l}_{\alpha,\beta}(n^{c},\mathbf{k})\langle \phi^{l^{\prime}}_{\alpha^{\prime},\beta^{\prime}}(A)|\frac{p_{x}}{m}|\phi^{l}_{\alpha,\beta}(A)\rangle \nonumber \\ +A^{\ast l^{\prime}}_{\alpha^{\prime},\beta^{\prime}}(n^{c},\mathbf{k})\times B^{l}_{\alpha,\beta}(n^{v},\mathbf{k})\langle \phi^{l^{\prime}}_{\alpha^{\prime},\beta^{\prime}}(A)|\frac{p_{x}}{m}|\phi^{l}_{\alpha,\beta}(B)\rangle \nonumber \\+B^{\ast l^{\prime}}_{\alpha^{\prime},\beta^{\prime}}(n^{c},\mathbf{k})\times A^{l}_{\alpha,\beta}(n^{v},\mathbf{k})\langle \phi^{l^{\prime}}_{\alpha^{\prime},\beta^{\prime}}(B)|\frac{p_{x}}{m}|\phi^{l}_{\alpha,\beta}(A)\rangle \nonumber \\+B^{\ast l^{\prime}}_{\alpha^{\prime},\beta^{\prime}}(n^{c},\mathbf{k})\times B^{l}_{\alpha,\beta}(n^{v},\mathbf{k})]
\langle \phi^{l^{\prime}}_{\alpha^{\prime},\beta^{\prime}}(B)|\frac{p_{x}}{m}|\phi^{l}_{\alpha,\beta}(B)\rangle\}.
\end{eqnarray}
In the gradient approximation, $\langle \phi^{l^{\prime}}_{\alpha^{\prime},\beta^{\prime}}|\frac{p_{x}}{m}|\phi^{l}_{\alpha,\beta}\rangle\cong\frac{\partial}{\partial k_{x}}\langle \phi^{l^{\prime}}_{\alpha^{\prime},\beta^{\prime}}|H|\phi^{l}_{\alpha,\beta}\rangle$, depending on the intralayer and interlayer hopping integrals combined with the Peierls phases. $M(n^{c},n^{v})$, which is determined by the relations among eight sub-envelope functions and the sublattice-dependent hopping integrals/Peierls phases, can account for the magneto-optical selection rules. By accurate calculations and  detailed examinations on the well-behaved LLs, the product of three terms in Eq. (4) remains the same or  changes the sign after the interchange of the initial and final states in the sublattices. Its direct summation leads to a finite (vanishing) velocity matrix element  for ${\Delta\,n=}$ odd (even) integers under the  $x$-polarization  (discussed latter in Fig. 3), and the opposite behavior is revealed in the $y$-polarization (Fig. 4).


\bigskip
\bigskip
\centerline {\textbf {III. RESULTS AND DISCUSSION}}%
\bigskip
\bigskip

The special lattice structure and complicated hopping integrals induce the rich band structures. Bilayer BP has a direct gap of ${E_g\sim\,1}$ eV near the $\Gamma$ point, as illustrated in  Fig. \ 1(d) by the black curves, being smaller than that (${\sim\,1.6}$ eV) of monolayer system. The highly anisotropic energy bands present the approximately linear and parabolic dispersions, respectively, along $\Gamma$X and $\Gamma$Y (${\hat k_x}$ and ${\hat k_y}$).  Energy gap is rapidly reduced by an electric field (blue curves). The semiconductor-semimetal transition appears at the critical strength ${E_{z,c}=0.32}$ (${V/\AA}$; red  curves), in which the valence and conduction bands are transformed into the linearly intersecting bands and oscillatory bands along $\Gamma$Y and $\Gamma$X, respectively. Two split Dirac-cone structures are situated on the right- and left-hand sides of the $\Gamma$ point (along ${+\hat k_y}$ and ${-\hat k_y}$). The extreme points remain at the $\Gamma$ point, accompanied with two saddle points on the opposite $k_x's$. The electronic states near the Dirac and $\Gamma$ points will be magnetically quantized into two distinct LL subgroups (discussed in Fig. 2) \cite{Wu2017}.

All the critical points and the constant-energy loops in the energy-wave-vector space will dominate the main features of LLs. At $E_{z}=0$, each LL is four-fold degenerate for each (${k_x,k_y}$) state in the presence of spin and localization-center degeneracies. The occupied LLs are asymmetric to the unoccupied ones about the Fermi level. The LL energy spectrum is very sensitive to the electric-field strength, as illustrated in Fig.\ 2(a). The conduction and valence LL energies, respectively, rapidly decline and grow in the increase of $E_z$. The lowest unoccupied and the highest occupied LLs, corresponding to the band-edge states, merge together at ${E_{z,c}}$, in which they are closely related to the magnetic quantization of electronic states near the Dirac points. A larger $E_z$ can extend the range of linear energy bands and thus double the degeneracy of low-lying LLs (from two neighboring Dirac points in Ref. \cite{Wu2017}). The LL anti-crossings and crossings occur frequently, since there are two competitive/cooperative LL subgroups initiated from the Dirac and $\Gamma$ points, respectively \cite{Wu2017}.

The amplitude, localization center and oscillation mode of the LL wave functions strongly depend on the electric-field strength. For bilayer BP systems, the eight sub-enevlope functions on the distinct sublattices have simple relations, so ${\Psi^{c,v}\,(A_1^1)}$ is illustrated to understand the critical characteristics and the $E_z$-dependence. In the absence of $E_z$, $\Psi^{c}(A^{1}_{1})=-\Psi^{c}(A^{1}_{2})=\Psi^{c}(A^{2}_{1})=-\Psi^{c}(A^{2}_{2})=\Psi^{c}(B^{2}_{2})=\Psi^{c}(B^{1}_{2})=-\Psi^{c}(B^{2}_{1})=-\Psi^{c}(B^{1}_{1})$ and $\Psi^{v}(A^{1}_{1})=\Psi^{v}(A^{1}_{2})=\Psi^{v}(A^{2}_{1})=\Psi^{v}(A^{2}_{2})=-\Psi^{v}(B^{2}_{2})=-\Psi^{v}(B^{1}_{2})=-\Psi^{v}(B^{2}_{1})=-\Psi^{v}(B^{1}_{1})$.
The quantum number $n^{c}$ ($n^{v}$) for each conduction (valence) LL are clearly identified from the number of zero points. For example, the four low-lying conduction/valence LLs have ${n^{c,v}=0, 1, 2; 3}$, as shown in Fig. 2(b). The localization centers are near the 1/2 and 2/2 positions of the enlarged unit cell of the crystal lattice, reflecting the magnetic quantization associated with the $\Gamma$ point. The former is sufficient in studying the magneto-optical properties. The well-behaved spatial distributions are similar to those in monolayer graphene \cite{SCChen2017}. In addition, the deeper/higher $n^{c,v}$ LLs might have  the side modes of $n^{c,v}\pm\,1, 2; 3$ due to the complicated interlayer hopping integrals.

A perpendicular electric field breaks mirror symmetry, leading to the probability transfer among the distinct sublattices and even the changes of spatial oscillation modes. With the gradual increase of $E_z$, the four sub-envelope functions on the first and second layers, respectively, have a simple relation, as revealed in the ${E_z=0}$ case (Figs. 2(b)-2(d)). Such functions change slowly for the conduction LLs, while  the opposite is true for the valence LLs. Their  oscillation modes  are dramatically changed from ${n^v}$ into ${n^v+1}$ as $E_z$ is close to the critical one, e.g., ${E_z=0.29}$ and 0.3 (Figs. 2(e) and 2(f)). And then, the ${n^c=0}$ and ${n^v=1}$ LLs are hybridized with each other, so that their wavefunctions present the combination of two oscillation modes  ((${E_z=0.32}$ in Fig. 2(g)). Furthermore, there exist two localization centers on the left- and right-hand sides of the 1/2 position, reflecting the existence of two neighboring Dirac-cone structures \cite{Wu2017}. The unusual/perturbed oscillation mode, with a main mode and significant side modes, are also revealed in other lowing-lying  LLs. In addition, the LL splitting will become obvious at deeper/higher energies, in which the $\Gamma$-induced  LL subgroup comes to exist and exhibits the anti-crossing behaviors with the Dirac-point-related one.  The various features of sub-envelope functions under distinct electric-field strengths imply the electrically tunable optical selection rules.

The magneto-optical excitations directly reflect the main features of magnetic quantization. The  available inter-LL excitation channels,  which arise from the transitions from the occupied ${n^v}$ states to the unoccupied ${n^c}$ ones, are denoted as ${(n^v,n^c)}$. They exhibit a lot of delta-function-like absorption peaks, being sensitive to external fields and polarization directions. For the parallel polarization direction, the magneto-absorption symmetric peaks, as shown in Fig. 3(a) at ${B_z=30}$ T,  are dominated by the selection rules of  ${\Delta\,n=1}$ and 3. The threshold peak comes from the (0,1) channel. There are many double-peak structures as a result of the asymmetric LL energy spectrum (Fig. 2(a)). Furthermore, the ${\Delta\,n=1}$  absorption structures (red circles) might be much higher than the  ${\Delta\,n=3}$ ones (black circles). For the former, a finite velocity matrix element is induced by the symmetric/anti-symmetric superpositions of eight sub-envelopes  in the initial and final LL states and the dominating intralayer hopping integral (${h^{11}_{12}}$) (Eq. 4)), according to the detailed analyses. The  extra selection rule, such as the latter, is driven by the side modes of the deeper/higher perturbed LLs under the interlayer hopping integrals. Specifically, all the inter-LL absorption peaks present the non-uniform intensities, regardless of the selection rules.

The complex relations between the Coulomb potential energies and the intrinsic interactions might dramatically change the magneto-optical selection rules. In addition to the ${\Delta\,n=1}$ excitation channels, the ${\Delta\,n=0}$ ones (green circles) come to exist quickly as $E_z$ gradually grows, as clearly indicated in Fig. 3(b) at ${E_z=0.1}$. These two kinds of absorption peaks appear alternatively, in which the neighboring two peaks of the former are merged together, and the threshold peak arises from the latter. The ${\Delta\,n=0}$ selection rule reflects the fact that an electric field could create an obvious asymmetry  between the valence and conduction sub-envelope amplitudes on the same sublattices (Fig. 2(b)), especially for the low-lying LLs. This asymmetry  also affects the peak intensities, so that the ${\Delta\,n=0}$ peaks are stronger (weaker) than the ${\Delta\,n=1}$ ones at lower (higher) frequency. The ${\Delta\,n=0}$ channels  will become the dominant excitations in the further increase of field strength, e.g., ${E_z=0.2}$ in Fig. 3(c), a result of the enhanced amplitude asymmetry (Fig. 2(c)). But when $E_z$ approaches to the critical field, the ${\Delta\,n=1}$ peaks grow rapidly, compared with the ${\Delta\,n=0}$ ones, e.g., $E_z$=0.29 in Fig. 3(d). This is attributed to the $E_z$-induced significant side modes in the low-lying valence LLs (Fig. 2(d)). Specially, for ${E_z\ge\,E_{z,c}}$, the Dirac-point-created LLs exhibit very strong absorption spectra at lower frequency, e.g., the prominent peaks of ${\Delta\,n=1, 0; 3}$ at ${\omega\,<0.06}$ eV under ${E_z=0.32}$ in Fig. 3(e). The main reason is that such LLs have the double degeneracy, and two localization centers with the main and side modes (Fig. 2(f)). Moreover, the $\Gamma$-related LLs also make important contributions to the higher-frequency absorption peaks.




The magneto-optical properties are strongly anisotropic even for the degenerate LLs. The spectral intensities  in the perpendicular polarization  greatly decline, as indicated from the comparison between Figs. 4(a) and 3(a). The $y$-polarization velocity matrix element  depends on the smaller hopping integrals, but is independent of the largest one (Eq. (4)). This illustrates the drastic changes of  peak intensities during the variation of polarization direction. The magneto-optical selection rules become ${\Delta\,n=2, 0; 4}$ (blue, pink; yellow circles; discussed in Eq. (4)). The ${\Delta\,n=1; 3}$ channels disappear under a specific relation  in the two-sublattice-dependent velocity matrix element. The ${\Delta\,n=2}$ and 0 channels, respectively, dominate absorption peaks at lower and higher frequencies. Apparently, the (0,2) channel creates the threshold peak. When the electric  field is below the critical one, only ${\Delta\,n=2}$ peaks could survive (Figs. 4(b)-4(d)). The ${\Delta\,n=0}$ selection rule is greatly suppressed through the strong asymmetry of LL amplitudes. However, more available excitation channels come to exist for ${E_z\ge\,E_{z,c}}$, e.g., the ${\Delta\,n=1, 0; 3}$ selection rules at ${E_z=0.32}$ in Fig. 4(e)). The similar phenomenon, the diversified selection rule, also appears  in the $x$-polarization (Fig. 3(e)), directly reflecting the characteristics of the Dirac-point-induced neighboring LL subgroups and the strong crossings/anti-crossings with the $\Gamma$-created LL subgroup.


The dependences of magneto-absorption spectra on the magnetic- and electric-field strengths deserve a closer examination. In the absence of  $E_z$, the peak frequencies, corresponding to the selection rules
(${\Delta\,n=1; 3}$ in Fig. 5(a) and ${\Delta\,n=2, 0; 4}$ in Fig. 5(b)), grow with the increasing $B_z$. The $B_z$-dependence deviates from the square-root and  linear relations,   being different from those in monolayer graphene and 2D electron gas.
 This is closely related to the unusual band structure (Fig. 1(d)). There exist certain important differences between the parallel and perpendicular polarizations in terms of the intensity, frequency and number of absorption peaks
The largest intralayer hooping integral involved in the $x$-polarization excitations is responsible for the  high anisotropy of absorption intensity. The threshold frequencies of the former and the latter  are, respectively, determined by
${\Delta\,n=1}$ and 2, so optical gaps are lower under the $x$-polarization. The monotonous $B_z$-depenence is also revealed in electric fields below the critical one.

On the other side, magneto-optical properties possess the  unique $B_z$-dependence beyond the critical electric field, as shown in Figs. 6(a) and 6(b) at ${E_z=0.32}$. Two categories of inter-LL excitations, which arise from the Dirac-cone- and $\Gamma$-valley-related LL subgroups (Fig. 2(a)), respectively, contribute to lower- and higher-frequency absorption peaks. The LL energies of the former and the latter, respectively, grow and decline in the increment of magnetic field, and so do their absorption peak frequencies. The $B_z$-dependent absorption frequencies become non-monotonous and abnormal when the anti-crossings of  two subgroups frequently appear  at sufficiently high magnetic field (${B_z>30}$ T). Specifically, the lower absorption frequencies due to the Dirac-cone LLs present the special $\sqrt {B_z}$-dependence for  ${B_z<30}$ T, as observed in monolayer graphene \cite{Plochocka2008,Orlita2015,JHHo2008}. The first absorption peak is induced by the occupied/unoccupied LL at the Fermi level and the nearest conduction/valence LL Its intensity is strongest among all the absorption peaks. The threshold  frequency is independent of  polarization direction; furthermore, it  increases with $B_z$ in the monotonous form.

The main features of the magneto-absorption spectra are very sensitive to the electric-field strength. For ${E_z<E_{z,c}}$, the frequencies of absorption peaks decrease rapidly  e.g., $E_z$ in Figs. 7(a) and 7(b) at ${B_z=30}$ T. On the other hand, the unusual magneto=optical properties are revealed in the range of ${E_z>E_{z,c}}$. They present the non-monotonous, oscillatory and discontinuous  $E_z$-dependences. Since the quantum modes are altered by a sufficiently high electric field, the new/original excitation channels will appear/disappear. Some absorption peaks could only survive in certain ranges of $E_z$'s. Obviously, the threshold channel is changed from (0,0) into (0,1) [(0,2) into (0,1)] near the critical electric field  under the $x$-polarization  (the $y$-polarization). This even leads to the abrupt change of threshold frequency.

The above-mentioned features of magneto-absorption spectra could be examined by various optical spectroscopies, such as, the infrared transmission \cite{Chuang2009,Orlita2009,Ubrig2011,Plochocka2012,Plochocka2008,Orlita2015,Zaric2004} and Raman scattering spectroscopies \cite{Berciaud2014,Henni2016}. Up to now, experimental measurements have confirmed the rich and diverse magneto-optical properties in graphene-related systems, e.g., Bernal graphite \cite{Chuang2009,Orlita2009,Ubrig2011,Plochocka2012}, AB-stacked few-layer graphenes \cite{Plochocka2008,Orlita2015}, and carbon nanotubes \cite{Zaric2004}. The low-lying absorption peaks in layered graphenes are identified to agree with the selection rule of ${\Delta\,n=1}$ \cite{Plochocka2008,Orlita2015}, in which the square-root and linear dependences of excitation frequencies on $B_z$, respectively, are attributed to the monolayer- and bilayer-like characteristics. The coexistent behavior is also observed in AB-stacked graphite \cite{Chuang2009,Orlita2009,Ubrig2011,Plochocka2012}. Moreover, the periodical  Aharonov-Bohm  effect is verified to exist  in cylindrical nanotube systems \cite{Zaric2004}. The magneto-optical properties in bilayer BP, the selection rule, number, frequency and intensity of absorption peaks, are greatly diversified/enriched by the polarization direction and external fields;  that is, they are easily tuned by the external factors. The experimental identifications are very useful in understanding the effects due to the geometric symmetry, intrinsic interactions; electric and magnetic fields. It should be noticed that Bilayer BP is very different from  AB- and AA-stacked graphenes \cite{Ho2010,Ho2010APL} with respect to the anisotropy/isotropy, selection rules; $B_z$- and $E_z$-dependences of magnet-optical spectra. In general, the latter present the almost isotropic excitations, a dominating selection rule of ${\Delta\,n=1}$, and the linear or square-root relation between $B_z$ and  frequency of absorption peak.



\bigskip
\bigskip
\centerline {\textbf {IV.\  Concluding Remarks}}%
\bigskip
\bigskip

The generalized tight-binding model, accompanied by the Kubo formula, is utilized to explore the rich and unique magneto-optical properties of bilayer BP. The main features of LLs, the field-dependent energy spectra and wavefunctions, are well characterized by the oscillation modes of sub-envelope functions on the distinct sulattices. They account for the selection rule, number, frequency and number of magneto-absorption peaks, strongly depending on polarization directions; electric and magnetic fields. The predicted results could be verified by magneto-optical spectroscopies, as has been done graphene-related systems \cite{Chuang2009,Orlita2009,Ubrig2011,Plochocka2012,Plochocka2008,Orlita2015,Zaric2004,Berciaud2014,Henni2016}.The theoretical framework could be further developed to fully understand the diverse magnetic quantization phenomena, especially for the generalization to emergent 2D systems covering few-layer silicene, germanene, tinene, antomonene, bismuthene, etc.

Bilayer BP exhibits the highly anisotropic magneto-absorption spectra, various selection rules, and the usual/abnormal field dependences. The spectral intensity declines obviously during the variation from the $x$- to $y$-polarizations. This is determined by whether the velocity matrix element is associated with the largest intralayer hopping integral. The selection rules, which come from the well-behaved LLs in the absence $E_z$, are, respectively, characterized by ${\Delta\,n=}$odd (even) integers for the $x$- and $y$-polarizations. The absorption frequencies deviate from the linear or square-root $B_z$-dependence. With the increase of $E_z$, the spectral intensities of available channels might be greatly enhanced/suppressed, or the extra selection rules come to exist. The non-monotonous and oscillatory field dependences are revealed in electric fields beyond the critical one, e.g., the drastic/dramatic changes in threshold frequency/channel near ${E_z,c}$. Such magneto-opical properties clearly illustrate the close relations among the geometric symmetry,  intralayer and interlayer atomic interactions, and external fields. There exist important differences between bilayer BP and graphene in the main features of magneto-optcail spectra.

\bigskip

\bigskip

\centerline {\textbf {ACKNOWLEDGMENT}}%

\bigskip

\bigskip

\noindent \textit{Acknowledgments.} This work was supported by the MOST of Taiwan, under
Grant No. MOST MOST 106-2112-M-022-001.

\newpage

\par\noindent ~~~~$^\star$e-mail address: yarst5@gmail.com

\par\noindent ~~~~$^\dag$e-mail address: szuchaochen@gmail.com

\par\noindent ~~~~$^\dag$$^\dag$e-mail address: ggumbs@hunter.cuny.edu

\newpage

\bigskip \vskip0.6 truecm

\noindent

\newpage

\begin{figure}[p]
\centering
\includegraphics[width=0.75\textwidth]{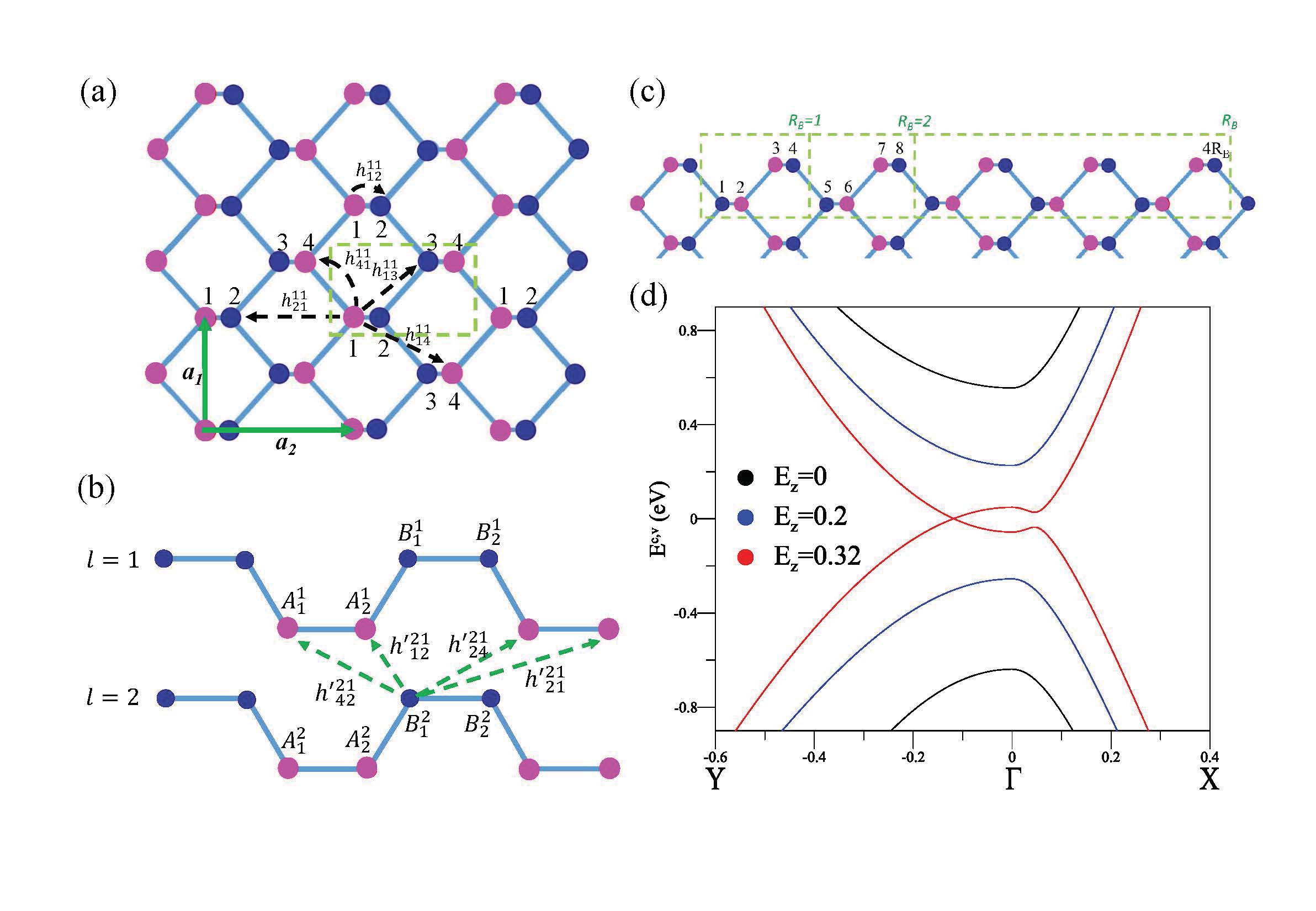}
\caption{Geometric structures of (a) monolayer and (b) bilayer phosphorus, respectively, for top and
side views with various atomic interactions. Also shown are (c) an enlarged unit cell in a uniform perpendicular magnetic field and (d) band structures under various electric fields.}
\label{figure:1}
\end{figure}

\begin{figure}[p]
\centering
\includegraphics[width=0.85\textwidth]{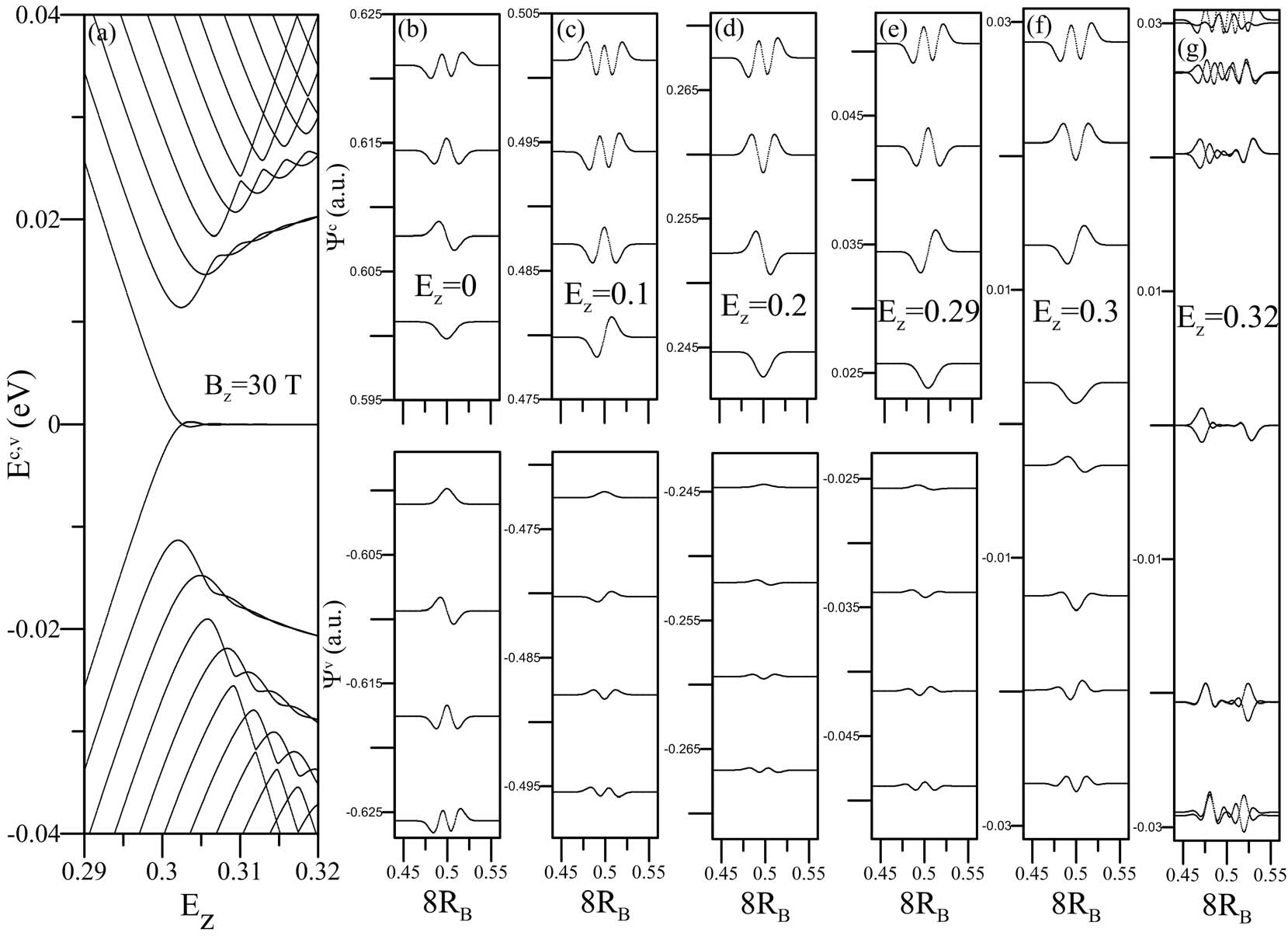}
\caption{The $E_z$-dependent LL spectrum at ${B_z=30}$ T (the left panel) and the corresponding
wavefunctions at various $E_z$'s (right panels).}
\label{figure:2}
\end{figure}

\begin{figure}[p]
\centering
\includegraphics[width=0.85\textwidth]{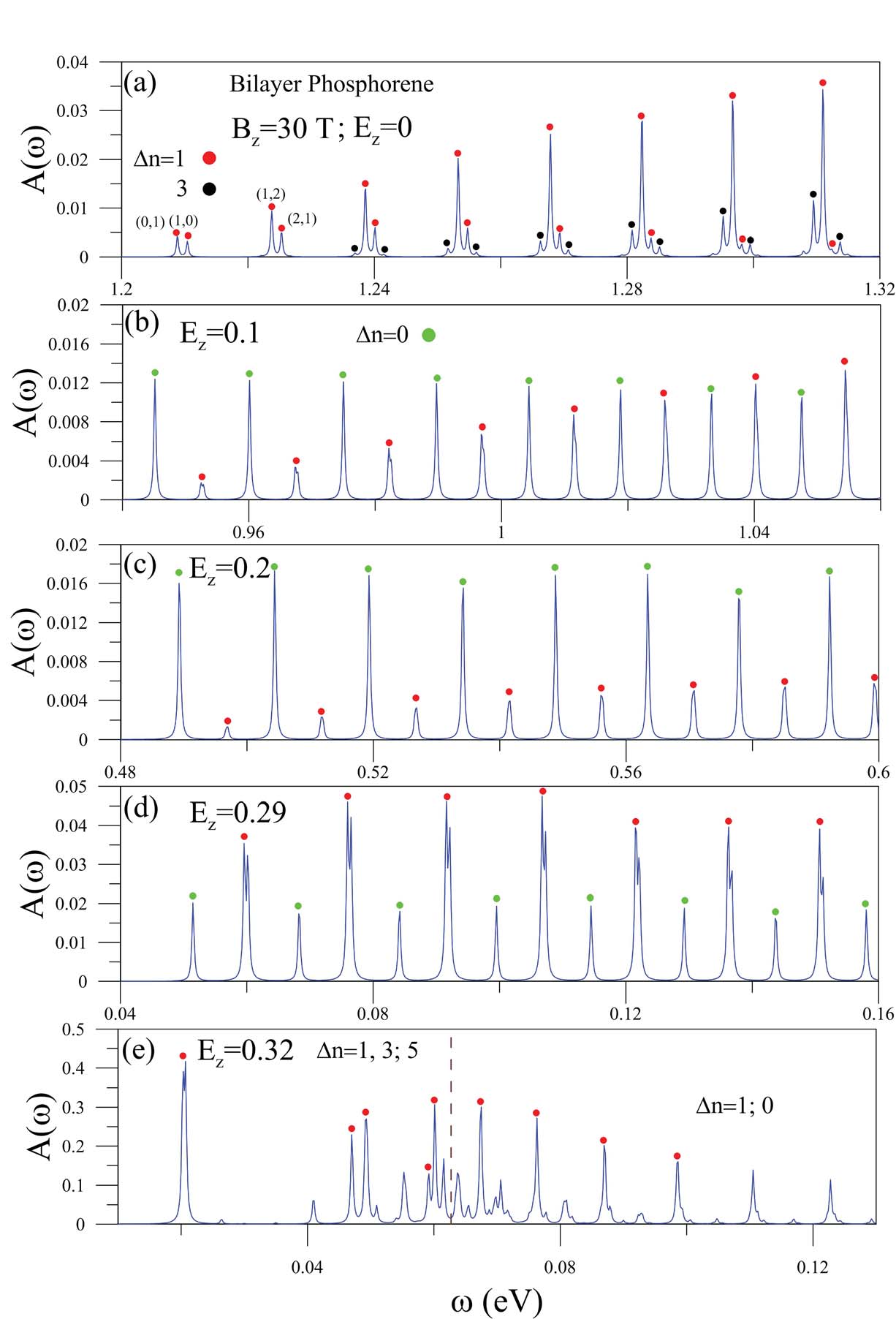}
\caption{The $x$-polarized magneto-optical absorption spectra at ${B_z=30}$ T for various electric-field strengths: (a) ${E_z=0}$, (b) 0.1, (c) 0.2, (d) 0.29 and (e) 0.32 (V/$\AA$).}
\label{figure:3}
\end{figure}

\begin{figure}[p]
\centering
\includegraphics[width=0.85\textwidth]{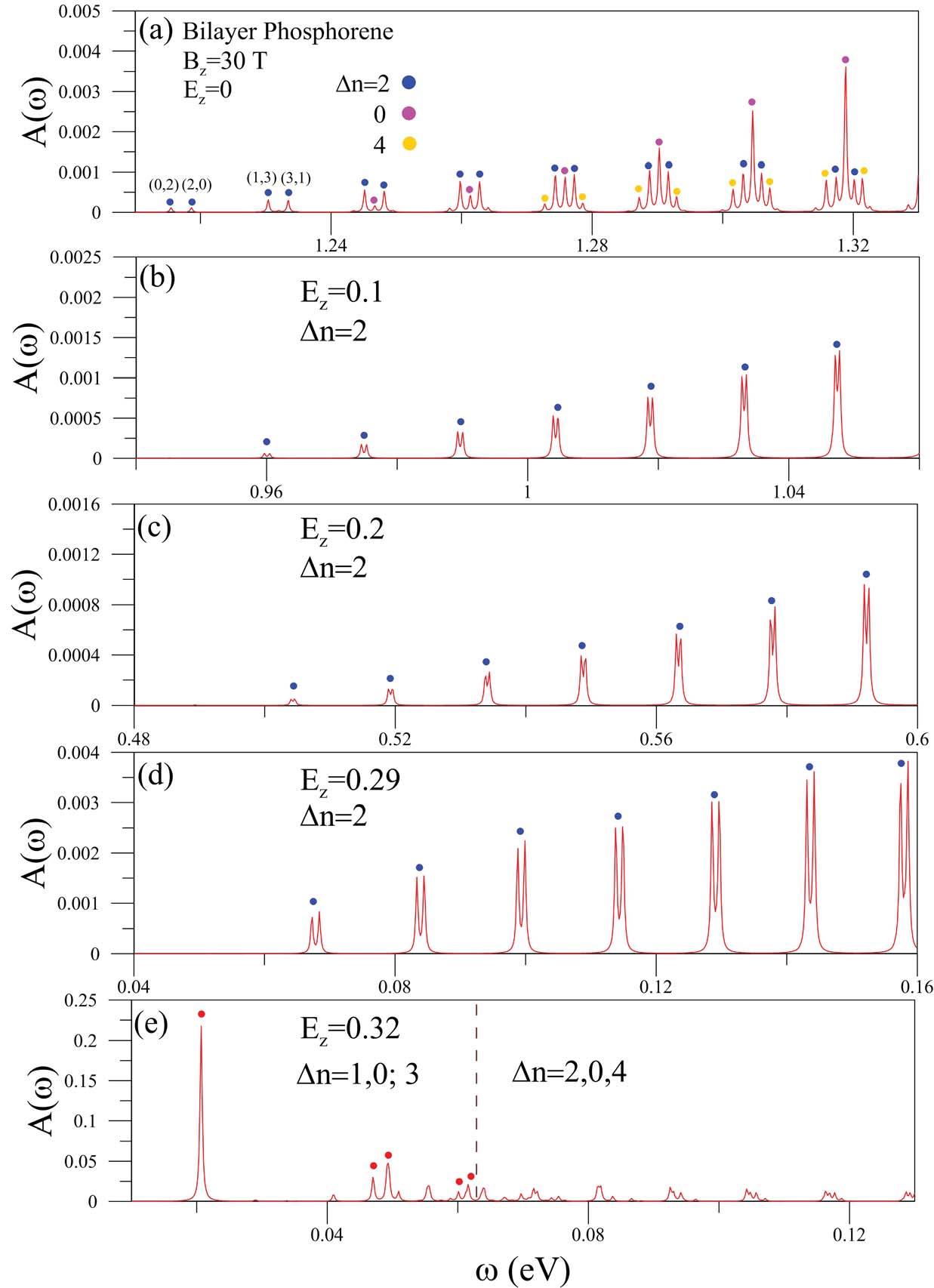}
\caption{Similar plot as Fig. 3, but calculated under the $y$-polarization.}
\label{figure:4}
\end{figure}

\begin{figure}[p]
\centering
\includegraphics[width=0.65\textwidth]{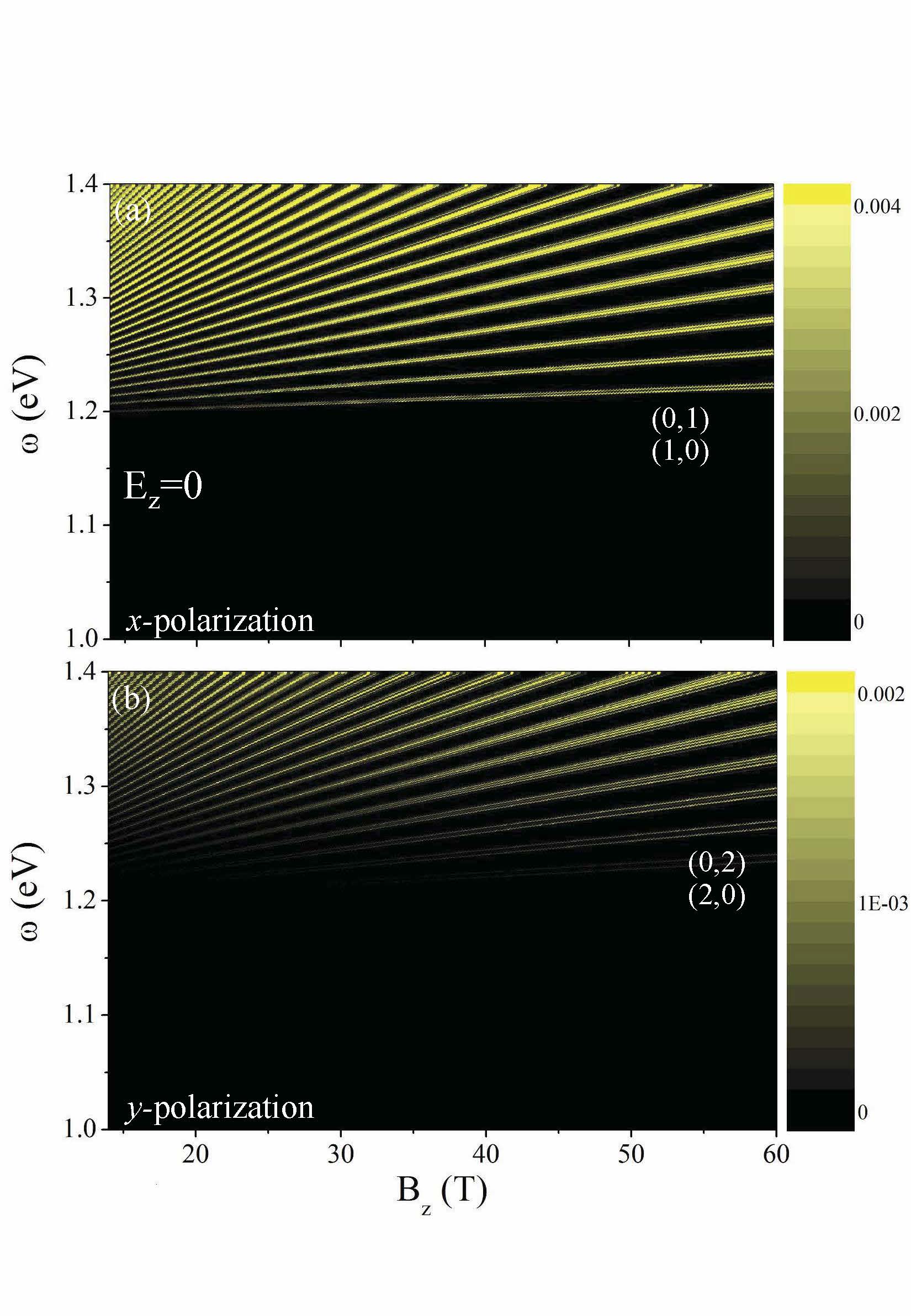}
\caption{Intensity plot of the $B_z$-dependent optical absorption spectrum at $E_z=0$ under (a) $x$- and (b) $y$-polarizations.}
\label{figure:5}
\end{figure}

\begin{figure}[p]
\centering
\includegraphics[width=0.85\textwidth]{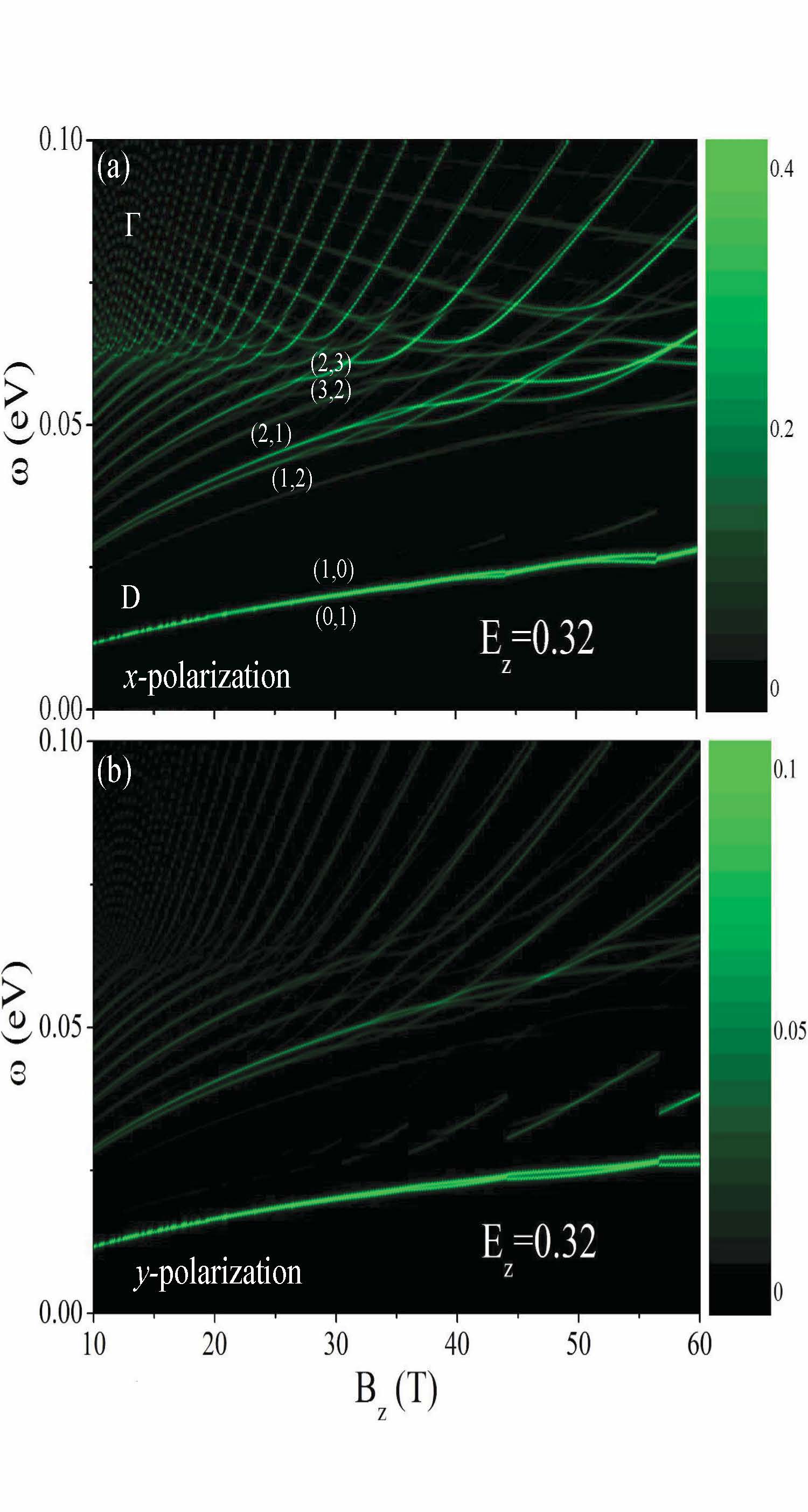}
\caption{Same plot as Fig. 5, but shown at ${E_z=0.32}$.
}
\label{figure:6}
\end{figure}

\begin{figure}[p]
\centering
\includegraphics[width=0.85\textwidth]{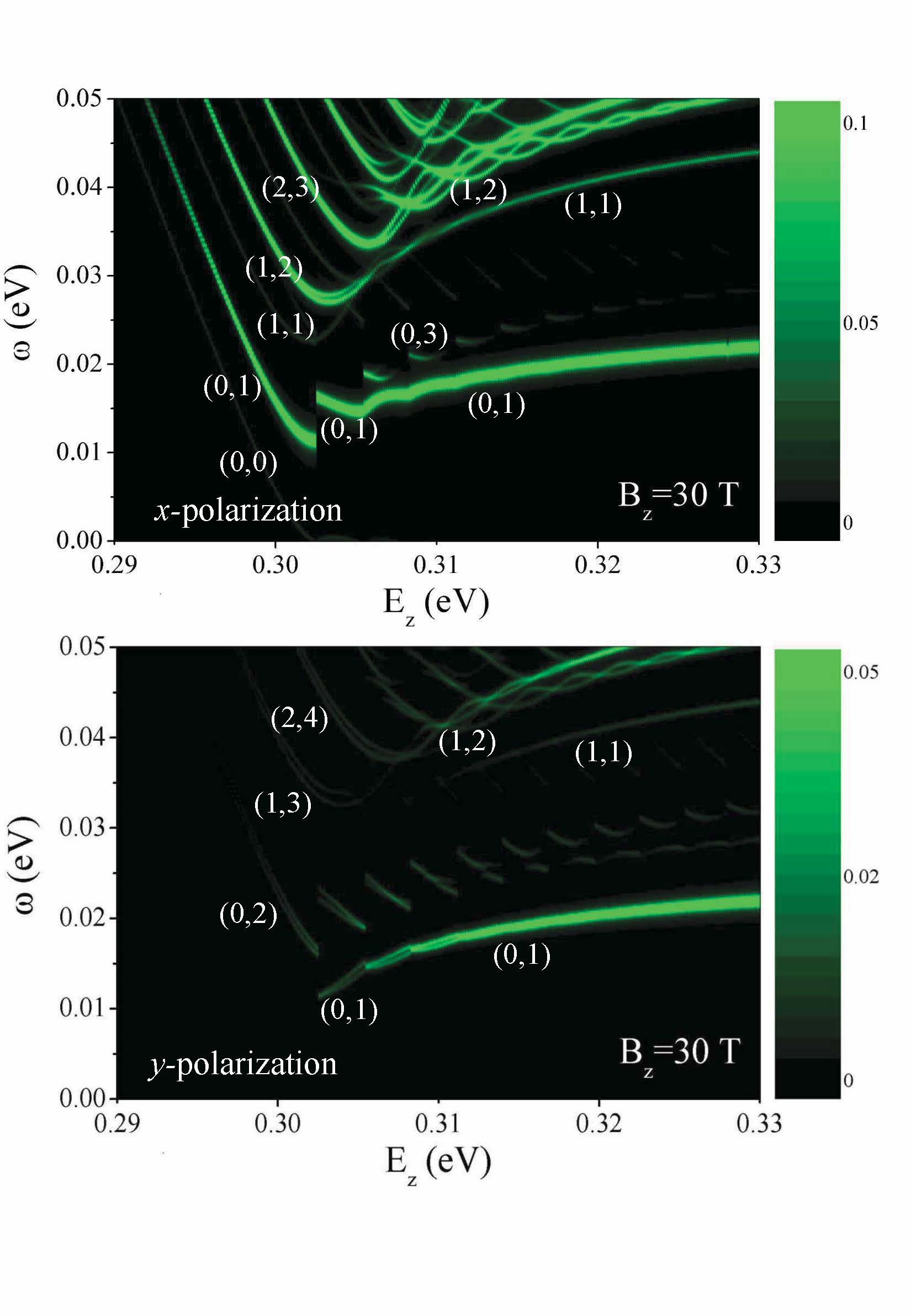}
\caption{The $E_z$-dependent absorption intensity at $B_z=30$ T under (a) $x$- and (b) $y$-polarizations.}
\label{figure:7}
\end{figure}

\end{document}